\documentstyle[12pt,epsf]{article}

\begin{document}

\newcommand{\sheptitle}
{Modified Dispersion Relations from Closed Strings in Toroidal Cosmology }
\newcommand{\shepauthor}
{Mar Bastero-Gil$^{(1)}$, Paul H. Frampton$^{(2)}$ and Laura Mersini$^{(1)}$} 
\newcommand{\shepaddress}
{$^{(1)}$ Scuola Normale Superiore and INFN, Piazza dei Cavalieri 7,\\ 
I-56126 Pisa, Italy \\
$^{(2)}$ Institute of Field Physics, Department of Physics and Astronomy,\\
University of North Carolina, Chapel Hill, NC 27599-3255, USA.}
\newcommand{\shepabstract}
{}

\begin{titlepage}
\begin{flushright}
hep-th/0110067\\
IFP-800-UNC\\
SNS-PH/01-08\\
\today
\end{flushright}
\vspace{.1in}
\begin{center}
{\large{\bf \sheptitle}}
\bigskip \medskip \\ \shepauthor \\ \mbox{} \\ {\it \shepaddress} \\
\vspace{.5in}

\bigskip \end{center} \setcounter{page}{0}
\shepabstract
\begin{abstract}
A long-standing problem of theoretical physics is the exceptionally small
value of the cosmological constant $\Lambda \sim 10^{-120}$ measured
in natural Planckian units. 
Here we derive this tiny number from a toroidal string cosmology
based on closed strings. In this picture the dark energy 
arises from the correlation between momentum and winding modes that
for short distances has an exponential fall-off with increasing values of the
momenta.The freeze-out by the expansion of the background universe for
these transplanckian modes may be interpreted as a frozen condensate
of the closed-string modes 
in the three non-compactified spatial dimensions. 
\end{abstract}

\vspace*{3cm}

\begin{flushleft}
\hspace*{0.9cm} \begin{tabular}{l} \\ \hline {\small Emails:
bastero@cibs.sns.it,
frampton@physics.unc.edu,
mersini@cibs.sns.it.} \\ 
\end{tabular}
\end{flushleft}

\end{titlepage}

\section{Introduction}

In this work we will attempt to make a quantitative argument 
about the origin of dark energy from string theory. 
The transition from string theory to conventional cosmology 
is of importance not only to theoretical physics in general but 
to inflationary cosmology in particular. Corrections to 
short distance physics due to the nonlocal nature of strings 
contribute to dark energy. The possibility to detect their
signature observationally  is very intriguing.  
In Ref. \cite{paper1} it was shown that a nonlinear dispersion function 
modifying the frequency of the transplanckian perturbation modes \cite{tp}
can produce  the right contribution to the dark energy of the universe
\cite{dark}. The  physics mechanism that gave rise to dark energy was
the freeze-out of these ultralow frequency modes by the expansion of
the background universe.  
Superstring duality \cite{tduality} was invoked to justify the dispersion 
function. This work attempts to carry out this derivation. 

In Section 2 we review some preliminaries of the
Friedman-Robertson-Walker (FRW) cosmological solutions   
found for string theory in a D-dimensional torus
\cite{vafa,mueller,tvafa,tseytlin}.
The quantum hamiltonian from closed string theory obtained in
\cite{tseytlin2} by using the correspondence principle between string
and quantum operators, is reviewed in Section 3. Although the
background is an FRW universe, it is globally nontrivial in
\cite{tseytlin2}, thus it allows two types of quantum string field
configurations, twisted and untwisted fields. 

Based on the equivalence between Euclidean path integral 
and statistical partition functions, we perform in Section 4 the
calculation of a  coarse-grained effective action \cite{blhu,huang}
for the momentum and 
winding modes of  the system described in Section 2 for the case of 3
expanding spatial dimensions $R$ in the $T^D$ toroidal
topology.  
The string scale is taken as the natural UV lattice cutoff scale of the
theory. The renormalization  
group equations (RGE) of the coupling constants for the winding and 
momentum modes describe the evolution from early to late times of 
their entanglement. 
Based on T-duality the whole spectrum is
obtained by exchanging momentum to winding modes
and $R \to R^{-1}$. Their coupling is strong when the radius of the 
torus $R$ is of the same order as the string scale $\sqrt{\alpha^\prime}$,
i.e. during  
the phase transition from a winding dominated universe to a 
momentum mode dominated universe. Due
to the expanding background, we have a non-equilibrium
dynamics  
and calculate the effective action by splitting our modes into the open
system degrees of freedom (low energy modes, mainly momentum modes) and the 
environment degrees of freedom (high energy modes, mainly winding modes).
The coarse-graining is performed by integrating out the environmental
degrees of freedom.  
The scale factor $a(t)=R(t)$ serves as the collective coordinate that 
describes the order parameter for the environment degrees of freedom. 
 The effective action calculated  
in this way contains the influence of the environment at all times in a 
systematic way and the coarse graining process encodes the dispersion 
function and corrections to short distance physics due to the 
correlation between the two types of modes in the system and
environment. This procedure results in the  
RGEs for the coupling constants that offers information about their 
running to trivial and nontrivial fixed points at early and late times, 
therefore the flow of one family of lagrangians (string theory phase) 
to another family of lagrangians at late times (conventional 3+1 quantum
theory). Results of this non-equilibrium phase transition are summarised in 
Section 5 with a discussion about the possibility of their observational 
signatures through the equation of state of the frozen short distance modes.
In this section we also briefly touch upon the issue of the two field
configurations in a globally nontrivial topology and the instabilities
in the theory arising from their interaction. A detailed summary of
the main coarse-graining formulas and procedure \cite{blhu} needed in
Section 4, 
are attached in the Appendix. In essence, the dark energy arises from the
study of the UV behavior of the correlations with environmental modes. 

\section{Toroidal String Cosmology.}

We consider the string cosmological scenario proposed by
Brandenberger and Vafa\footnote{Herein referred to as the BV model.}
\cite{vafa,tvafa,others}.   
Strings propagate in compact space, a box with 
D spatial dimensions and periodic boundary conditions, the $T^D$ torus.
It was argued that \cite{vafa} a thermodynamic description  of the
strings with positive specific heat, is well defined only when all the
spatial dimensions 
are compact.

Let us begin with the Universe placed in a $T^D$ box
with a size of the order of the string scale, that we are taking to
be the Planck scale. In such a 
space, string states also contain winding modes, which are
characteristic of having an extended object like a string, 
``winding'' around the compact spatial dimension, besides the usual
momentum modes, and oscillator modes with energy independent of the 
size of the box. The
energy of the winding modes 
increases with the size of the box as $w R$, while the energy of the
momentum modes decreases as $m/R$. The spectrum is symmetric under the
exchange $R \leftrightarrow 1/R$ and $m 
\leftrightarrow w$. This symmetry known as T-duality \cite{tduality} is not
only a symmetry of the spectrum but of the theory. 

The BV model \cite{vafa} argues that if the Universe expands
adiabatically in more than 3 spatial dimensions, it would 
not be possible to maintain the winding modes in thermal
equilibrium. As their energy density grows with the radius,  
their number would have to decrease, for example through annihilation
processes. But typically strings do not meet in more than 3 spatial
dimensions  and do not interact with each other; therefore the 
winding modes fall out of equilibrium \cite{numerical}. In
summary, their growing energy density will tend to slow down the expansion
of the universe and eventually stop it.  
But if the Universe starts to contract, the dual
scenario of the momentum modes opposing contraction would take place
and the Universe 
may oscillate between expanding/contracting eras. In what follows we
use this argument of \cite{vafa} to justify the assumption that only
$D=3$ dimensions of the $T^D$ torus will expand to create an FRW
universe.

Cosmological solutions for an arbitrary number of anisotropic
toroidal spatial dimensions $T^D$ were found by Mueller in
\cite{mueller}. He studied the cosmology of bosonic strings
propagating in the background defined by a time-dependent dilaton field,
$\Phi(t)$, and space-time metric 
\begin{equation} 
ds^2_d=G_{\mu\nu}(X) dX^\mu dX^\nu=-dt^2+ \sum_{i=1}^D 4 \pi R_i^2(t)dX_i^2\,,
\label{metric}
\end{equation}
The radii of the torus, $R_i(t)$, become the time-dependent
scale-factors, and the spacetime dimensions is $d=1+D$. The equations
of motion of the bosonic string in 
background fields are obtained from the following action\footnote{The
antisymmetric tensor field is taken to be zero.} \cite{action}
\begin{equation} 
I= \frac{1}{4 \pi \alpha^\prime} \int d^2 \sigma \sqrt{g} \left[ 
g^{mn} G_{\mu\nu} (X) \partial_m X^\mu \partial_n X^{\nu} + \frac{1}{2}
\alpha^\prime \Phi R^{(2)} \right] \,,
\end{equation} 
where $g_{mn}$ is the two-dimensional world-sheet metric, and
$R^{(2)}$ the world-sheet scalar curvature. The background field
equations are obtained by imposing the condition that the theory be free from
Weyl anomalies. To lowest order in perturbation theory this leads
to the equations:   
\begin{eqnarray}
\beta_{\mu\nu}^G &=& R_{\mu\nu} + \nabla_\mu \nabla_\nu \Phi =0 \,,\\
\beta^\Phi &=& \frac{d-26}{3 \alpha^\prime} - R + (\nabla \Phi)^2 -2
\nabla^2 \Phi =0 \,,
\end{eqnarray}  
Using the metric given in Eq. (\ref{metric}), they reduce to:
\begin{eqnarray} 
\ddot{\Phi} - \sum{i} \frac{\ddot{R}_i}{R_i} &=&0 \,, \label{bg1}\\
 \frac{\ddot{R}_i}{R_i} + \sum_{j\neq i} \frac{\dot{R}_i
 \dot{R}_j}{R_i R_j} - \dot{\Phi} \frac{\dot{R}_i}{R_i} &=&0 \,,\\
\ddot{\Phi} - \frac{1}{2} \dot{\Phi}^2 + \sum_{i<j} \frac{\dot{R}_i
 \dot{R}_j}{R_i R_j} &=& \frac{ d-26}{3 \alpha^\prime} \,. \label{bg3} 
\end{eqnarray} 
When $D=25$, the solutions obtained in \cite{mueller} are: 
\begin{eqnarray}
e^{-\Phi(t)} &\propto& t^p \,, \\
R_i(t) &\propto& t^{p_i} \,, 
\end{eqnarray}  
with the constraints,
\begin{equation}
\sum_{i=1}^D p_i^2 = 1 \,,\;\;\;\;\;
\sum_{i=1}^D p_i = 1 - p \,. 
\label{muellers}
\end{equation}
Note that these solutions are found in the absence of matter sources. 
In general the backreaction of the matter action of the strings 
in $T^D$ alters the solutions for the background geometry\footnote{See
\cite{tvafa,tseytlin}  
and references therein for the geometry solutions in the presence of a
matter action. Inclusion of matter 
sources alters the solutions of \cite{mueller} 
due to the backreaction of the winding modes, such that 
the scale factor approaches asymptotically a constant 
value at late times
.}. It is clear that we can have an arbitrary number of compact
spatial dimensions 
$D_c$ with $p_i < 0$, that are decreasing with time
\footnote{We do not address the concern that the time dependence of
the compactified $R_i$  
endangers the constancy of the dimensionless parameters in the $D=3$
theory.}, and $D-D_c$ expanding spatial 
dimensions with $p_i >0$. Among the many solutions found in
\cite{mueller} we select the solution $D-D_C=3$  that although it is
not unique it is justified by the BV argument. 
The assumption that our Universe is expanding in only 3 spatial
dimensions, with the remaining $D-3$ being small and compact, as well
as considering a constant dilaton field\footnote{The authors of \cite{tvafa}
argued that a constant dilaton background may not be consistent with
a $high$ temperature phase of strings thermodynamics.} 
($p=0$), are consistent with Mueller's solutions
Eqs. (\ref{muellers}). The issue of stabilising the dilaton is beyond
the scope of this 
paper, and we assume that the dilaton has acquired a mass
and become stable at some fixed value. It is also assumed 
that the backreaction of the matter string sources 
on the backround geometry is small enough such that the deviations from the 
FRW metric, Eq. (\ref{muellers}), can be neglected.

Due to the toroidal string cosmology, the three expanding dimensions
contain both types of modes: $momentum$ and $winding$, propagating 
in the 3+1 FRW space-time. The
number of winding modes at each 
stage of the evolution of the Universe is determined by the
dynamics of the background. In the next section, we touch base with
quantum field theory through correspondence principle between string
and quantum operators, in order to use coarse
graining techniques for studying the influence of the winding modes
on the momentum modes as the Universe expands. 

\section{Quantum Hamiltonian from Closed String Theory. }

Let us consider BV model \cite{vafa} of a 
D-dimensional anisotropic torus with radius $\bar{R_i}$, by including
the dynamics of both modes: momentum modes, $p_{1,i}=m/\bar R_i$
(where $m$ is the 
wavenumber), and winding modes with momenta $p_{2,i}=w\bar
R_i/\alpha^\prime$. The 
dimensionless quantity for the radius is
$R_i=\bar{R_i}/\sqrt{\alpha^\prime}$, where $\alpha^\prime$ is 
the string scale. Based on the arguments reviewed in Section 2, we
{\it choose} a 
cosmology with three toroidal radii equal and large $R \gg 1$ in
units of the string or Planckian scale, with 
the other $(D - 3)$
toroidal radii equal and small $R_C \ll 1$. Here the subscript $C$
refers to compactified dimensions.  
Then,  $R(t)$ becomes the scale factor
for the 3+1 metric in conventional FRW (Friedman Robertson Walker)
cosmology $R(t)=a(t)$, while $R_C$ corresponds to
the radius, in this factorizable metric, of the $D-3$ compact
dimensions $z_j$ that decrease with time,
\begin{equation}
d s_D^2 = -dt^2 + 4 \pi R^2(t) dx_i^2 + 4 \pi R^2_{C}(t) dz_j^2 =
a(\eta)^2 [- d\eta^2 + dy^2] + d s_{D-3}^2\,. 
\label{metric2}
\end{equation}
Using the string
toroidal solution of \cite{mueller} the time-dependence of these radii is:
\begin{equation} 
R(t) = \alpha_U t^{p_U}
\end{equation}
\begin{equation}
R_C(t) = \alpha_C t^{p_C}
\end{equation}
The solutions in Ref. \cite{mueller} show that $p_U$ and $p_C$ depend
on the dimensionality $D$ in an interesting way. There is a plethora
of possible solutions but if we assume, for example, that the dilaton
is time-independent and the compactification is isotropic we find that
for $4 \leq D < \infty$, then $0.5 \leq p_U < 1/\sqrt{3} \simeq 
0.577$. Let us 
take $D=4$ where the scale factor behaves as a radiation-dominated
universe; if, in fact, $D \geq 5$ we can assume that the $D-4$
additional dimensions have $p_C^{'} \ll p_C$ to achieve the same
result. In this case, $p_C = -0.5$. Here we do not, however, need to
specialise to a particular solution.

What we have in mind for the dark energy is the correlation of momentum to
{\it winding} string modes. 
The question is, given the well-known form for the kinetic energy of these
strings, {\it e.g.} \cite{kikkawa}, how to describe best the interaction
between the winding and momentum modes. Some aspects are addressed in
\cite{kikkawa} who focuses on the smallness of temperature $(T/T_H)$.
For temperature $T$ very much below the
Hagedorn or string temperature $T_H$ we expect  that
only very small winding numbers $w_i = 0$ or $1$ in the compact
dimensions are of any significance \cite{kikkawa}. Similar arguments
apply to the momentum modes $m_i$ for the time-reversed case.  

Let us consider the small parameter $\delta(t)$, taken to be:
\begin{equation}
\delta = \frac{R_C}{R} \sim t^{p_C - p_U}
\end{equation}
For the case $D=3$ (d=4), for example $\delta \sim t^{-1} \sim (T/T_H)^2$
and is an extremely small number ($\sim 10^{-60}$) at present. The
point is that in the $\delta \rightarrow 0$ limit these modes are
in separate spaces and for very small $\delta$ are therefore expected
to be highly restricted. The compactified dimensions can be integrated
out, and we are left with the momentum and winding modes in the
remaining $D=3$ spatial dimensions.

The partition function for this system was  calculated, from first
principles, by summing up over their momenta in \cite{kikkawa}:
\begin{equation}
Z= \sum_\sigma e^{-n_\sigma \epsilon_\sigma}\,,
\end{equation} 
where $n_{\sigma}$ is the number of strings in 
state $\sigma$ with energy $\epsilon_{\sigma}$
\begin{equation}
\epsilon_\sigma=p_0=\sqrt{\left(\frac{m}{R}\right)^2 + (w R)^2 + N +
\tilde{N} -2 } \,,
\label{epsilon}
\end{equation}
and $\sigma$ counts over $(m,\,w)$, with the constraint $N-\tilde{N}=m
w$ for closed strings where $N$ and $\tilde{N}$ are the
sums over the left- and right- mover string excitations, respectively.
By now, in Eq.(\ref{epsilon}),
we are considering only the large 3 spatial dimensions.
The string state can also be described by 
its left and right momenta, $k_L = p_1+p_2$, $k_R=p_1 - p_2$. The string
state for left and right modes can be  
expanded in terms of the creation and annihilation operators
$\alpha_m$, $\tilde\alpha_n$, with higher excitation string states
given by $N=\sum_{n=1}^{\infty} \alpha_{-n}\alpha_n$ (similarly
for $\tilde{N}$),  and string energy 
$L_0 + \tilde{L}_0 = p_1^2 +p_2^2 + (N +\tilde{N} -2)/\alpha^\prime$.   

We would like to write the
path integral for this configuration in terms of quantum
fields\footnote{Below we use quantum string equations under the assumption 
that the dilaton is massive and stable.}. The path integral is
calculated from the hamiltonian density.   
In order to use the correspondence between the Euclidean path integral
of the persistence vacuum amplitude $|\langle in | out \rangle|^2$ and
the partition function $Z$, we need to write a hamiltonian density over the
fields in configuration space  in such a way that its Fourier transform
in $k$-space corresponds to the string energy expression Eq. (\ref{epsilon}). 

Thus in writing a Coarse-Grained Effective Action (CGEA), the kinetic terms
are unambiguous while for the interaction terms we must appeal to simplicity
and the requirement of T-duality. Closed-string field theory
 provides guidance, since in {\it e.g.} \cite{belopolsky} truncation
at a quartic coupling can be sensible, and this will lead to a CGEA
which is renormalisable and satisfies useful RG equations. 

Generally, closed string field theory contains couplings of all non-polynomial
orders. In a semi-classical approximation
we may restrict to genus $g=0$ since the genus $g$ contribution
is proportional to $\hbar^g$ \cite{SB}.

The quantum hamiltonian is in any case known for the 
classical string in axi-symmetric or toroidal
backgrounds \cite{tseytlin2}. They explicitly
calculated the quantum hamiltonian and demonstrated the correspondence
principle between the string operators $L_0$, $\widetilde L_0$ and
quantum field operators in the form (in the notation of \cite{tseytlin2})
\begin{eqnarray} 
\hat{H}&=&\hat{L}_0+ \hat{\widetilde{L}}_0 = \frac{1}{2}\alpha^\prime
\left( -E^2 + p_a^2 + \frac{1}{2} (Q_+^2 +Q_-^2)\right) + N + \widetilde{N} -2
c_0  \nonumber \\
& & -\alpha^\prime \left[ (q+\beta) Q_+ + \beta E \right]
-\alpha^\prime \left[ (q-\alpha) Q_- + \alpha E \right] J_L \nonumber
\\
& & \frac{1}{2} \alpha^\prime q \left[(q+2 \beta) J_R^2 + (q-2 \alpha)
J_L^2 + 2 (q+\beta -\alpha) J_R J_L \right] 
\label{tseytlin}
\end{eqnarray} 
\begin{equation}
\hat{L}_0- \hat{\widetilde{L}}_0 = N- \widetilde{N} -mw
\end{equation}
where $J_{R,L}$ are bilinear quadratic operators in terms of creation
and annihilation operators and the higher string 
oscillators $N, \tilde N$ contribute the string mass. Therefore the
$J_R^2$ term would be a quartic interaction in terms of creation and
annihilation operators.

This particular solution is for a cylindrical
topology (Melvin model) where the uncompactified
$x_1$ and $x_2$ are written in polar coordinates
$x_1 + ix_2 = \rho e^{i \phi}$ and
$x_3$ is also uncompactified (but could be compactified along
with additional similar coordinates), together with time and one
additional compactified dimension $y \subset (0, 2 \pi R)$.
Although an exact solution for the hamiltonian of the string 
matter in a toroidal background is not yet known, a 
quartic potential energy was advocated and found in \cite{tseytlin} by  
arguments similar to those of Eq. (\ref{tseytlin}), for the classical
string and the three string coupling level.  
We take this as an indication, in the subsequent
section (if the exact solution were known to all orders), 
that an quantum hamiltonian analogous to Eq.(\ref{tseytlin})
for closed strings
on a torus, similarly containing only
quartic terms as suggested by \cite{tseytlin2}, exists 
for our present case 
of $(T_3) \times (T_{D-3}) \times (time)$ and focus on the 
uncompactified 3 spatial dimensions.

The hamiltonian depicted in Eq.(\ref{tseytlin}) is for a static 
background,i.e a constant scale factor
 $R(t)$. In the next section, we base our calculation in the 
coarse-grained effective action (CGEA) formalism where the 
dynamics  of an expanding background is replaced by scaling on a
static background. 

Thus Eq.(\ref{tseytlin}) which applies to a static background (as in Eq.(\ref{epsilon}))
can be generalized to a cosmologically-expanding background as in
Eq.(\ref{metric2}) by using this technique of re-scaling, as we shall
discuss in the next Section. This strategy is necessitated by the absence
of an exact string solution in the time-dependent background.

\bigskip
\bigskip

\section{Coarse Grained Effective Action (CGEA) and RGE's} 

\subsection{General case of the d=D+1 Universe}

Our system of winding and momentum modes is described by nonequilibrium
dynamics due to the expanding background spacetime. All the information about
the evolution of these 
modes will be contained in the effective action. 
Therefore we need to write the path integral in the configuration
space of the quantum fields in order to obtain the effective
action. This information must be extracted from the torus analogs of
the quantum mechanics hamiltonian of Eq.(\ref{tseytlin}) such that its
Fourier transform  in momentum $k$-space recovers the
string energy spectrum     
Eq. (\ref{epsilon}). Correlation functions are obtained by using the
correspondence between the Euclidean path integral of the persistence
vacuum amplitude $|\langle in | out \rangle|^2$ and the partition
function $Z$.  
All the string quantum operators below are promoted to quantum field
operators with the corresponding hamiltonian density ${\cal H}(t,x)$ in
configuration space derived from the quantum string hamiltonian
${\cal H}(t)$. 

The following calculations are done in the conformally flat 
background Eq. (\ref{metric2})\footnote{We do not address 
issues of the matter backreaction on the geometry. 
They are treated in \cite{tvafa,tseytlin}.}, through the scaling of the 
fields and operators with the conformal factor $a(\eta)$.
The momentum field $\phi_1(R,x)$ and the winding field
$\phi_2(R,x)$ are defined by the relation:
\begin{equation}
\phi_i(x)= \int e^{i p_i x} \phi_i(p_i) d^3p_i\,, \;\;\;\int |\nabla
\phi_i|^2 d^3x= \int d^3p_i p_i^2 \phi_i(p_i) \,,  
\end{equation}  
where
\begin{equation}
\nabla= R \partial/\partial x \ = \partial/\partial y \,,
\end{equation} 
and $p_i=p_1,\,p_2$. 
Let us also define two new fields, $\psi_L(R,x)$ and $\psi_R(R,x)$,
with momenta $k_L,\,k_R$ that are the left and right combinations of the
Kaluza Klein momentum and winding modes 
\begin{eqnarray}
\psi_L(R,x) &=& \phi_1(R,x) + \phi_2(R,x) \,, \label{psi1}\\
\psi_R(R,x) &=& \phi_1(R,x) - \phi_2(R,x) \,. \label{psi2}
\end{eqnarray}
These fields live in the expanding (3+1) spacetime
dimensions. Similarly there is another set of fields $\Psi_{c,a}$ that are
functions of the compact dimensions $z_a$. Their energy
contribution to the total hamiltonian density is ${\cal H}_C(p_a) = A_a p_a^2$
where $A_a$ is a constant parameter with dimensions of inverse volume
of the compact space, and $p_a$ are the momenta of 
these fields in the extra compact dimensions, with $a$ running over the
$D-3$ dimensions. 

The Hamiltonian density ansatz that would describe the energy of our 
two string states in the $D=3$ expanding dimensions 
with energy $H = L_0 +\widetilde L_0$, 
including the 
oscillators from string's higher excitations $(N +\widetilde N
-2)/\alpha^\prime$,    
is similar to the hamiltonian of spin waves in a periodic
lattice\footnote{Torus is obtained by identifying the first and the  
last lattice sites, thus the periodicity.}. The Ginzburg-Landau
hamiltonian for a Heisenberg magnet 
obtained in \cite{tseytlin2} by means of CFT bears similarity with
$\lambda \phi^4$ quantum field theory in a well-known
manner.Our lattice spacing is given by the string scale 
$\sqrt{\alpha'}$. Therefore the hamiltonian density can be written
for this dual lattice in terms of wave functional ``spin'' fields
$\psi_L(R,x),\psi_R(R,x)$ 
of Eqs. (\ref{psi1}), (\ref{psi2}) as follows
\begin{equation}
{\cal H}= {\cal H}_3 + {\cal H}_C \,,
\end{equation}
with
\begin{equation} 
{\cal H}_3 = |\nabla \psi_L|^2 + |\nabla \psi_R|^2 |+ |\nabla
\psi_L||\nabla \psi_R|+ m_0^2 ( |\psi_L|^2 + 
|\psi_R|^2) + g_1 (|\psi_L|^4 + |\psi_R|^4) + g_2 |\psi_L|^2
|\psi_R|^2 \,,
\end{equation}
where the fields $\psi_L,\,\psi_R$ are expanded in terms of the mode
functions $u_n,\tilde u_n$, 
\begin{equation}
\psi_L = \Sigma u_n b_n + u_n^* b_n^+\,, \;\;\;\; \psi_R = \Sigma
\tilde u_n \tilde b_n + \tilde u_n^* \tilde b_n^+\,,
\end{equation}
and  $b_n,\tilde b_n$ are the normalised
quantum creation and annihilation operators of 
$\alpha_n, \tilde \alpha_n$. 
The commutation relation for the unnormalised operators are such that
$[\alpha_n, \alpha_m^+] = \omega_{\pm}\delta_{nm}$ with 
$\omega_{\pm}$ the frequency of left and right moving modes. 

The periodic lattice condition $N-\widetilde N=mw$ introduces  
an interaction term in the hamiltonian ${\cal H}_3$ of 
the form $\nabla\psi_L\nabla \psi_R$. 
In terms of the 2-component state $\Psi_{EN}
= (\psi_L, \psi_R)$, the hamiltonian  reads,
\begin{equation}
{\cal H}_3= |\nabla \Psi_{EN}|^2 + \nabla \Psi_{EN} \widehat{X} \nabla
\Psi_{EN} +  
m_0^2 |\Psi_{EN}|^2 +
g_1 |\Psi_{EN}|^4  + (g_2 - 2 g_1) |\Psi_{EN} \widehat{X}
\Psi_{EN}|^2 \,, 
\end{equation}
with 
\begin{equation} 
\widehat{X}= \left( \begin{array}{cc} 0 & 1/2 \\ 1/2 & 0 \end{array}
\right) 
\end{equation}
The system is known as the dual momentum-space lattice, and for $g_2=2
g_1$ reduces 
to the XYZ model of condensed matter. Let us for simplicity limit to
the XYZ model case, $g_2=2 g_1$, for the rest of this paper. 

These periodic lattice systems studied in 3+1 dimensions in terms of Bloch 
wavefunctions have a solution which respect to lattice translation 
invariance, $exp(-p l)$, with the lattice spacing ``$l$''  
equal to the string scale $\sqrt{\alpha^\prime}$. 
The interaction term,
in the tight-binding approximation, lifts the degeneracy between  
the energy eigenstates due to the leakage/tunnelling of the
wavefunction from one lattice site to the neighbour site. 
As a result the gap energy produced between
the ground (bound) state and higher excitation states is 
\begin{equation}
p^2 \Delta_p = p^2 |\cos(2 \theta)|= p^2 ~ |2 cos^2 (\theta) - 1| \,,
\label{gap}
\end{equation}
in which 
\begin{equation}
p l =p \sqrt{\alpha^\prime}= \sqrt{\alpha^\prime(p_1^2 + p_2^2)}=
\sqrt{(\frac{m}{R})^2 + (wR)^2} \,,
\end{equation} 
and $\theta \rightarrow \theta + i p l$. Therefore, 
\begin{equation} 
\Delta_p \leq 2 \cosh^2(p l) -1 \,. 
\label{massgap}
\end{equation}
The first term in $\theta$
is a pure phase of rotation of the ``spin-wave'' in the dual lattice,
but the second term describes the tunnelling of the wavefunction to the
nearest neighbour\footnote{ In condensed matter this is known as
Coulomb dipole type of vortex interaction.}.
The gap energy of Eq. (\ref{gap}) introduces a correction to the
kinetic energy, such that in momentum space the hamiltonian reads
\begin{equation} 
{\cal H}_3= z_p p^2 |\Psi_{EN}|^2 + m_0^2 |\Psi_{EN}|^2 + g_1
|\Psi_{EN}|^4 \,,
\end{equation} 
with $z_p = 1 + \Delta_p$. This correction
contributes to the wavefunction renormalization constant of the field
$\Psi_{EN}$. We can then make a partial (finite)
renormalization of the hamiltonian in order to recover the canonically
normalised kinetic term, such that
\begin{eqnarray} 
\Psi_{EN} &\rightarrow& \widetilde{\Psi}_{EN}=z^{1/2}_p \Psi_{EN} \,,\\
m_0^2 &\rightarrow& \tilde{m}^2_0 = z_p^{-1} m_0^2 = \frac{m_0^2}{1 +
\Delta_p} \,, \label{mass}\\
g_1 &\rightarrow& \tilde{g}_1 = z_p^{-2} g_1 \,.  
\end{eqnarray} 
The hamiltonian density ${\cal H}_3$ finally reads, 
\begin{equation} 
{\cal H}_3 = |\nabla \widetilde{\Psi}_{EN}|^2 + \tilde{m}^2_0
|\widetilde{\Psi}_{EN}|^2 + \tilde{g}_1 
|\widetilde{\Psi}_{EN}|^4 \,.  
\label{hamiltonian} 
\end{equation} 

The action in D-dimension therefore is
\begin{eqnarray} 
S_D &= & \int dt~R_Cd^{(D-3)} z_a~ a(t) d^3 x \left({\cal
H}_3[\widetilde{\Psi}_{EN}(x)]+ {\cal 
H}_C[\Psi_{c,a}(z)] \right)  \\
&=& V_C \int a(t) dt~d^3 x {\cal H}_3[\widetilde{\Psi}_{EN}(x)]+ V_U \int R_C
dt~d^{D-3}z {\cal 
H}_C[\Psi_{c,a}(z)] \,,
\end{eqnarray}
where $V_C$ ($V_U$) is the volume factor obtained from integrating out the
contribution from the compact $(D-3)$  (uncompact) dimensions.
The partition function $Z$ is then,
\begin{eqnarray}
Z &=& Z_c Z_3 \,, \\
Z_c &=& 
\int D \Psi_{c,a} e^{-V_U \int R_C dt~d^{D-3} z {\cal
H}_C[\Psi_{c,a}]} \,, \\ 
Z_3 &=& \int D
\widetilde{\Psi}_{EN} e^{- V_C \int a(t)dt~d^3 x {\cal
H}_3[\widetilde{\Psi}_{EN}]} \,. 
\end{eqnarray} 
The contribution  $Z_c$ to the path integral is easy to
calculate since the integral over the compact dimension fields is a
simple gaussian, 
\begin{equation}
Z_c = \int D \Psi_{c,a} e^ {-V_U \int d^{D-3} p_a A_a \Psi_{c,a} p_a^2
\Psi_{c,a}}
= \prod_a \sqrt{\frac{\pi}{A_a V_U}} \,.
\end{equation}
The contribution of these fields to the path integral is proportional to
the volume of the compact space $R^a_C$ they live in, thus their
contribution is relevant only around the string scale because at late
times the volume of the compact metric decreases rapidly with time. In
either case their contribution rescales the normalisation constant of
the path integral, which we will allow for the moment to be
arbitrary\footnote{Varying the action with respect to the metric care
should be taken to account for the effect of the compact metric volume
on the Newtons constant of the reduced (3+1) metric.}, such that
\begin{eqnarray}   
Z &=& N^\prime \int D
\widetilde{\Psi}_{EN} e^{- V_C \int a(t) dt~ d^3 x {\cal
H}_3[\widetilde{\Psi}_{EN}]} \\ 
&=& N^\prime Z_3 \,.
\end{eqnarray} 
The volume of the compact dimensions, $V_C$, is
roughly of order unity in terms of string units, and it can be
reabsorbed into the parameters of ${\cal H}_3$. 

We would like to find a simplified description for the dynamics of our
non-equilibrium system, consisting of both winding and momentum
modes, while incorporating the backreaction of the short wavelength
modes to it, in the reduced $(3+1)$ dimensions. This is done by
carrying out the necessary steps of coarse graining which are the
following \cite{blhu}: 1) distinguish the system from the environment,
2) coarse 
grain the environment, 3) measure how the coarse grained environment
influences the system in providing an effective dynamics for our
reduced system. As we will see below the environment, consisting of the
short wavelength modes, has a time dependent order parameter due to
the expanding background universe, thus the need for using
non-equilibrium dynamics methods\footnote{A whole program with a
detailed treatment of the conceptual and formal techniques of
coarse-graining has been pioneered and developed by Hu and
collaborators in \cite{blhu}. They showed that for a special
class of expansion, the dynamics of spacetime can be equivalently
replaced by a scaling transformation with time playing the role of the
scaling parameter.}.  

\subsection{The D=3 Universe}

At this point, in evaluating the reduced 3 dimensional partition
function $Z$, we want to 
separate our modes into system (S) + environment (E) degrees of freedom, and
coarse grain by integrating out the degrees of freedom for the
environment. This amounts to finding out the backreaction of the
coarse grained environment on the system, and
eventually leads to the RGE's\footnote{The running of the coupling
constants with time depends on how one selects the
environment and the system.}. We use many of the results and the
approach of \cite{blhu} in what follows. We will consider as
environment all the short wavelength 
modes with momenta 
\begin{equation}
(E): \;\;\;\;\; \frac{\Lambda}{b} < p^E
=\frac{1}{\sqrt{\alpha^\prime}}[(m/R)^2 + (w R)^2]^{1/2} < \Lambda \,, 
\label{env}
\end{equation} 
where the cutoff $\Lambda= (\alpha^\prime)^{-1/2}$ is the string
scale because $(\alpha^\prime)^{1/2}$ is identified with the lattice spacing
$l$, and $b=a(t)/a(t_0)$ is the coarse grain scaling parameter, where 
$t_0$ is the initial time. The scale factor $a(t)$ plays the 
role of the collective coordinate describing the environmental degrees
of freedom. Time in this procedure is playing the role of a scaling
parameter and dynamics is being replaced by scaling \cite{blhu}. This is an
artificial procedure (known as Kadanoff-Migdal transform
\cite{kadanoff}) that relates 
the microscopic and macroscopic properties of a system based on the
existence of scaling properties of the system in the infrared
limit. Thus $a(t)$ is treated simply as a parameter while carrying out
the coarse graining in this ``static limit'', 3-dimensional Minkowski
field theory of the foliation $a(t)$=constant  hypersurface at each
$t_n=t_0+n \Delta t$.  

The system modes are the ones with:
\begin{equation}
(S): \;\;\;\;\; p^S< \frac{\Lambda}{b} \,, 
\label{system} 
\end{equation}
>From the above definitions of system and environment,
Eqs. (\ref{system}) and  (\ref{env}), 
at initial times when $b\approx 1$ we have $p \le \Lambda/b$ thus all
our modes, momentum and winding, are in the system;   
but at later times when $b \geq 1$ more and more winding modes
systematically transfer  
to the environment because the condition of
Eq. (\ref{system}), $p=\frac{1}{\sqrt{\alpha^\prime}}[(m/R)^2 + (w
R)^2]^{1/2} \le 
\Lambda/b$ is satisfied only for vanishingly small winding numbers
$w \rightarrow 0$. As $t$ becomes large, the system contains
$m \leq R \Lambda$, $w=0$, i.e.  
all the modes except $m \leq R \Lambda, w =0$ have transfered to the environment. The 
Euclidean path-integral of this 2 field system
$\widetilde{\Psi}_{EN}$  with 
hamiltonian density given in Eq. (\ref{hamiltonian}), is\footnote{From
now on, we drop the subindex ``3'' from the the action.}  
\begin{equation}
Z= | \langle R<1 | R >1 \rangle |^2 = \int \prod_{\Lambda/b \leq p
\leq \Lambda} D\tilde \Psi_E \prod_{0 \leq p \leq \Lambda/b} D\tilde 
\Psi_S~ e^{-S[\widetilde{\Psi}_{EN}]} \,,
\label{partition}
\end{equation}
where the field is split into high and low energy as follows:
$\widetilde{\Psi}_{EN}= \tilde \Psi_S + \tilde \Psi_E$, e.g. $\tilde
\Psi_E$ denotes the modes 
 with 'environment' momenta $p^E$ given by
Eq. (\ref{env}). After this splitting of the modes 
into (System+ Environment), we can separate the terms in the action
$S[\widetilde{\Psi}_{EN}]$ into:
\begin{equation}
S[\widetilde{\Psi}_{EN}]=S_S[\tilde \Psi_S]+S_0[\tilde
\Psi_E]+S_I[\tilde \Psi_E, \tilde \Psi_S]\,,  
\end{equation} 
where $S_S,S_0$ are the action depending on system, 
environment variables and $S_I$ is the piece that depends on 
the interaction of system variables to the environment variables.
\begin{eqnarray} 
S_S[\tilde \Psi_S] &=& \int a(t)dt \int d^3 x ( \tilde \Psi_S G_S^{-1} \tilde \Psi_S + g_1
\tilde \Psi_S^4)\,, \\ 
S_0[\tilde \Psi_E] &=& \int a(t)dt \int d^3 x \tilde \Psi_E G_E^{-1} \tilde \Psi_E \,, \\
S_I[\tilde \Psi_E, \tilde \Psi_S] &=& \int a(t) dt \int d^3x g_1 \left[
4 \tilde \Psi_S^3 \tilde \Psi_E + 6 \tilde \Psi_S^2 
\tilde \Psi_E^2 + 4 \tilde \Psi_S \tilde \Psi_E^3 + \tilde \Psi_E^4
\right] \,. 
\end{eqnarray}
$G_{S,E}$ are Green's functions for open system ($S$) and environment ($E$)
given by:
\begin{eqnarray}
G_S^{<\Lambda/b}[p^S< \Lambda/b]&=& [ (p^S)^2 + \tilde{m}_0^2]^{-1} \,, \\
G_E^{\Lambda/b}[p^E \leq \Lambda/b]&=& [ (p^E)^2 + \tilde{m}_0^2]^{-1} \,,
\end{eqnarray}
The Green function for the whole (closed system, S+E) system $G[p]$ satisfies:
\begin{equation}
G[p]= G_S^{<\Lambda/b}[p^S] + G_E^{\Lambda/b}[p^E]\,.
\end{equation}
After integrating out the high energy $p^E$ modes  
Eq. (\ref{env}) in the action, we are left with an effective action
that depends only on the system variables $p^S < \Lambda/b$, such that:
\begin{equation}
S_{eff}[\tilde \Psi_S] = S_{S}[\tilde \Psi_S] + \Delta S[\tilde \Psi_S] \,.
\label{action}
\end{equation}
$S_{S}[\tilde \Psi_S]$ is the portion of
the action that all along depends only on the system variables.   
The term  $\Delta S$ results from the interaction of
the system with the environment, but it depends only on the system 
variables after the coarse-graining. It gives rise to corrections 
$\delta \tilde{m}_0^2$ and $\delta \tilde{g}_1$ to
the mass and coupling parameters in the action (see \cite{blhu}, Appendix
for details). Therefore, 
the effective action $S_{eff}$ will have the same form as
$S_S[\tilde \Psi_S]$ with parameters $\tilde{m}^2$ and $\tilde{g}$ defined as 
\begin{eqnarray} 
\tilde{m}^2= \tilde{m}_0^2 + \delta
\tilde{m}_0^2 \, \\
\tilde{g}= \tilde{g}_1 + \delta
\tilde{g}_1 \,.
\end{eqnarray}
We assumed that 
$\cosh^2(p\sqrt{\alpha^\prime})$ in the expression of $\Delta_p$ that
enters in the mass term $\tilde{m}_0^2$ Eq. (\ref{mass}), is a slowly
varying function of 
momenta $p$ and consider it to be a constant while carrying out the
procedure of coarse graining\footnote{We are keeping only first order
corrections to the mass term due to the $\Delta_p$. Contributions from
higher order terms to the mass correction $\delta \tilde{m}_0^2$, 
like $\Delta_p^\prime$, $\Delta_p^{\prime \prime}$,...have been ignored.}. 

Let us rescale our variables in the effective  action
Eq. (\ref{action}), such that 
\begin{equation}
p^\prime = b p,\;\;\;\; \widetilde{\Psi}^\prime(p^\prime)=
b^{-(D+2)/2}\widetilde{\Psi}_{EN}(p^\prime/b) \,.  
\end{equation}
Clearly, the original cutoff $\Lambda$ and range of momenta are
restored after rescaling. Dynamics has been replaced by scaling of
parameters in a  {\it static spacetime} \cite{blhu}. 
This procedure can be
repeated n times, for very small time increments $\Delta t= 
(t_{f}-t_0)/n$ between the initial and final times,
\begin{eqnarray}
S_{eff}(\tilde \Psi^\prime) &=& b^{-D} \int d^D p^\prime 
\tilde \Psi(p^\prime/b)
\left [\left(\frac{p^\prime}{b}\right)^2 + \tilde{m}^{ 2} +
\tilde{g} \langle \tilde \Psi^2
\rangle 
\right] \tilde \Psi(p^\prime/b) \\
& = & \int d^D p^\prime \tilde \Psi^\prime(p^\prime)
\left[ \left(p^\prime \right)^2 + b^2 \tilde{m}^2 + b^{4-D} \tilde{g}
\langle \tilde \Psi^2 \rangle 
\right] \tilde \Psi^\prime (p^\prime) \,. 
\end{eqnarray}
$S_{eff}[\tilde \Psi^\prime]$ will have the same form as the original one in
Eq. (\ref{action}) provided that we identify the mass term and
coupling constant
\begin{equation}
\tilde{m}^{\prime 2}= b^2 \tilde{m}^2 \,,\;\;\;\;
\tilde{g}^\prime=b^{4-D} \tilde{g} \,.
\end{equation}
Although we are formally keeping the dimensionality to be  an arbitrary
$D$ in discussing the RGEs below, in fact our reduced system has $D=3$
and we take that limit at the end. 
Repeating this procedure n times (with $n \rightarrow \infty$),
results in the RGEs for the coupling constants.

The canonical two-point correlation function at high energy for 
system-environment interaction is calculated from the path 
integral of the canonical fields $\tilde{\Psi}_{S,E}$ in
momentum space (Fourier transform of $G^{\Lambda/b}$). 
It is related to the correlation function of the original fields $\Psi_E$ 
(which decreases at high energy) as follows 
\begin{equation}
\langle \Psi_E \Psi_E \rangle= 
\frac{\langle \tilde \Psi_E \tilde \Psi_E \rangle}{z_p} 
\,. 
\label{corrfunct}
\end{equation}
where $z_p = 1 + \Delta_p$ and $\Delta_p$ is given in Eq.(\ref{massgap}).
This is the crucial result for the 
interpretation of the cosmological dark energy.
Because of the mass gap, the correlation function is suppressed exponentially
in p-space. It is very familiar that a mass gap leads to an exponential
fall off in x-space, but here for the dual lattice the exponential fall off
is in momentum space. This may be traced to the T-duality of the closed 
strings and the resultant interchange of the IR/UV limits.
As we will show, for the rescaled fields $\tilde{\Psi}_E$ the correlation function
increases at high energy leading to an exponential decrease in 
the dispersion $\omega(p)$.

The two-point correlation function at low
energies, $p^S \leq \Lambda/b$ is related to the canonical one 
in the same way. The canonical two-point function is the Fourier transform of
$G^{<\Lambda/b}$, and at low momenta it goes like a 
polynomial: 
\begin{equation}
\langle \tilde \Psi_S \tilde \Psi_S \rangle= \frac{1}{(p^S)^2 + \tilde{m}^{\prime 2 }}
\,. 
\end{equation}
The reason for this different behaviour of the correlation function is
because before splitting $\Psi$ into high + low energy modes, the
total canonical two-point correlation function is: 
\begin{equation}
\langle \tilde{\Psi^\prime} \tilde{\Psi^\prime} \rangle = G[p^\prime]=
\frac{1}{p^{\prime 2} + \tilde{m}^{\prime 2}} 
\,,
\label{correlation}  
\end{equation}
When writing it in terms of the two point function of the 
original fields (by dividing with the $z_p$ normalisations factor), 
it goes as an exponential for large momentum $p$, and as a
polynomial for low momenta.

The correlation length is given by
\begin{equation}
\xi\simeq (\tilde{m}^\prime)^{-1}\,.
\label{corrlength}  
\end{equation}
Clearly, $\xi$ diverges at very high momenta (early times) that 
indicate that the correlation length is very large and signals a 
phase transition. Alternatively, the correlation length goes to zero at 
low momenta (late times) indicating that the theory becomes local at late times.

Let us denote $\tau=\ln b$, and $\epsilon=4-D$. The RGEs for this
system at hand are known from the analogy of our partition function,
\begin{equation}
Z = N \prod_{p^\prime < \Lambda} \int D \tilde{\Psi^\prime}
e^{-S_{eff}[ \tilde{\Psi^\prime}]} \,, 
\label{partition2}
\end{equation}
to the dual lattice Ising model,
\begin{eqnarray}
\frac{d g_1}{d \tau}&=& \epsilon g_1 -A ( 36 g_1^2 + g_2^2) \,, \label{dg1}\\
\frac{d g_2}{d \tau}&=& \epsilon g_2 -A ( 24 g_1 g_2 + 8 g_2^2) \,,
\label{dg2}\\ 
\frac{d x}{d \tau} &=& 2 x + 12 (1-x) (1 + \frac{g_2}{6 g_1})
\frac{g_1}{\Lambda^\epsilon} \,, \label{dx}
\end{eqnarray}
with $x= \tilde{m}^{\prime 2} \Lambda^{-2}$, and $A$ a numerical constant. 

For the case we considered, $g_2=2 g_1$ ($g_1 \rightarrow \tilde{g}$)
thus Eq. (\ref{dg1}) reduces and becomes 
identical to Eq. (\ref{dg2}). The solution to the RGEs will tell us
the running of the couplings 
constant $\tilde{m}^{\prime 2}$ and $\tilde{g}^{\prime}$ to their
nontrivial fixed points with time, 
$\tilde{m}^{\prime 2}=f_m(b)$, $\tilde{g}^\prime= f_g(b)$. These
relation $f_i(b)$ are replaced in 
the expression for the correlation function and length,
Eqs. (\ref{correlation}) and (\ref{corrlength}). Note that
$G^{-1}[p]$ is the dispersed frequency with short distance
modifications contained in the $\tilde{m}_0^2$ term:
\begin{equation}
\tilde{m_0}^2 \stackrel{p \to \infty}{\rightarrow}
\frac{m_0^2}{2 \cosh^2 p \sqrt{\alpha^{\prime}}} \simeq \frac{1}{2}
m_0^2 e^{-2 \sqrt{\alpha^{\prime}} p} \,. 
\label{result}
\end{equation}
where $b_f =a(t_f)/a(t_0)$ and $t_f$ is the final time that can be
taken to be future infinity. 
It is clear from the expression for the modified
canonical mass of Eq.(\ref{result}),
which originated from the interaction between the system
and the environment at short distances,
that the correlation length of Eq.(\ref{corrlength})
diverges exponentially at high energies and the correlation
function between the original fields falls off
exponentially, Eq.(\ref{corrfunct}).
 From the RGEs, the correlation length $\xi$ and the quartic coupling
constant $g_1$ vanish with time, $\xi 
\rightarrow_{t\rightarrow \infty} 0$, and the system is dominated only
by $free$ momentum modes $(m/R)$ ($ w \rightarrow 0$). The 
RGEs in combination with the CGEA thus describe the dynamics 
evolution of our entangled system of winding and momentum modes at 
early times to a free gas of momentum modes at late times, 
due to the backreaction of the high energy environmental (mainly
winding) modes.  
Also notice that by using the Tolman relation for the temperature in
an expanding background, $T= T_c/a(t)$, we can express 
the RGEs and the partition function $Z$ as functions of 
temperature rather than the scale factor $a(t)$ (or equivalently $b$), i.e.: 
\begin{equation}
Z=Z[ \frac{T-T_c}{T_c}]    \,,
\end{equation}
where $T_c$ corresponds to $a(t_0)$, i.e. $b=1$. The correlation
length $\xi$ diverges around $b=1, T=T_c$, $\xi
\stackrel{t \rightarrow 0}{\rightarrow} \infty$. Thus our system breaks
down due to strong correlations at early times or high temperatures,
but this simply signals the Hagedorn 
phase-transition at around the critical temperature $T_c$. 

\section{Dark Energy from Closed String Theory. Discussion}

We argue that closed strings on a toroidal cosmology 
lead to a plausible explanation of the dark
energy phenomenon. Although bosonic strings have been used,
it is expected that superstrings will lead to a similar conclusion.
Certainly it is crucial that closed strings are involved because
open strings do not have the same aspect of winding around the torus.

The scale factor of the universe $a(\eta)$ has been used 
as a collective coordinate for the environment
degrees of freedom, and as the fundamental scaling parameter in the
coarse-graining. 
The choice of a $D=3$ expanding cosmology was chosen phenomenologically. An
argument for this choice in the BV model was presented in 
\cite{vafa}, and we believe this argument does provide a possible
justification. It is encouraging that 
inclusion of branes gives a similar result \cite{others}. 
It has further been assumed that the mass gap $\Delta_p$ can be
safely assumed to be slowly-varying during
our coarse-graining procedure.

We would like to make the reader aware of another subtlety related to the
torus topology of our background. A globally nontrivial topology like 
$T^3 \times R^1$ admits two types of quantum field configurations,
twisted and untwisted fields, due
to the periodic and anti-periodic boundary conditions imposed
on the fibre bundle of the manifold. This is a long-standing
problem \cite{ford} that does not have a definite remedy. The problem is
the following: twisted fields can have a negative
two-point function. These fields
interact with each other while preserving the symmetries of the hamiltonian.
Their interaction thus contributes a negative mass squared term
to the effective mass of the untwisted field due to
the negative two-point function of the twisted field and render the
untwisted field unstable. It is often assumed that Nature simple chose 
to preserve the untwisted configuration only or forbids their interaction
due to some, as yet unknown, symmetry \cite{laura}. 
 
String theory preserves
Lorentz invariance. This symmetry has been broken for the 
open system of our low energy string modes due to the 
backreaction from the coarse grained environment. 
Their correlation results in our dispersion relation.
If a specific frame must be chosen, it 
could be {\it e.g.} the rest frame of the CMB.
The formalism needed for the calculation of the stress-energy
tensor and the equation of state of the non-linearly dispersed short-distance
modes in the presence of Lorentz non-invariance and an expanding background
while lacking an effective lagrangian for this short-distance 
physics requires further development.
The initial condition for our model is a vacuum state conformally equivalent 
to the Minkowki spacetime - the so-called Bunch-Davis vacuum\cite{BD}.
Finally, before summarising we should note that if there are other
modes without the exponential suppression at high $k$, all that we
need is one such mode to lead to the frozen tail comprising the dark energy.

The high wave number behaviour $e^{-ak}$ of
the dispersion relation $\omega(k)$ leads again to the
correct estimate for the dark energy as a fraction $\sim 10^{-120}$ of
the total energy during inflation. 
This dark energy is certainly completely stringy because
our derivation depends 
on the existence of winding modes, as seen by the role of
the generalised level-matching condition
\[
N - \widetilde{N} = \Sigma_{i}m_iw_i
\]
This correlation between momentum and winding modes leads to the
quantum hamiltonian 
\footnote{We would like to remind the reader of the
approximation made in Sec.3 for obtaining the toroidal string quantum
hamiltonian since an exact solution for this class of backgrounds does
not exist as yet.} 
and 
hence to the interpretation of the dark energy
as the weak correlation with the winding mode energy at short distances.
\footnote{The correlations in the transplanckian regime contribute to
the total
energy density of the universe with two types of modes: The excitation
modes of these correlations with energy less
than the current Hubble expansion rate, $H$, are currently frozen by the
expanding background therefore their kinetic energy is nearly zero. All
the other modes in the transplanckian regime whose frequency is higher
than H oscillate and redshift away at a rate that will be determined by
their equation of state.}
The excitation modes of these correlations with energy less 
than the current expansion rate are currently frozen by the 
expanding background. 

Within string cosmology there has always been the question of the
fate of the winding modes in the uncompactified three spatial dimensions,
whether they combine to a single string per horizon which 
wraps around the universe. Our
remedy is intuitively appealing that while the momentum
modes are in evidence as quarks, leptons, gauge bosons, etc. 
the winding modes are now
condensed uniformly in the environmental background, 
hence with a weak correlation at short distances to the momentum modes, 
frozen by the expansion of the FRW universe in the form of the dark energy.

The observed small value $\Lambda \sim 10^{-120}$ in natural units
has an explanation in the toroidal cosmology of closed strings and thus the
dark energy provides an exciting opportunity to connect string theory to
precision cosmology.
We may argue that numerically the
size of the cosmological constant in the present approach
is a combination of the string scale
and the Hubble expansion rate in the sense that
$\Lambda/M_{Planck}^4 \simeq 10^{-120} \simeq (H_0/M_{Planck})^2$.
Therefore the correct amount
of dark energy obtained by this frequency dispersion
function does not require any fine tuning and relies,
besides a physical mechanism (such as freeze-out),
only on the string scale as the parameter of the theory.
However, our approach does not solve the second puzzle about the dark
energy namely, the coincidence problem for the following reason: The
expansion rate of the universe is determined by the {\em total energy
density} in the universe by the relation given in the Friedman equation.
As can be seen from our dispersion function which approaches conventional
cosmology in the subplanckian regime ($k \le M_{pl}$), most of the other
contributions to the energy density are not frozen modes. Therefore the
Hubble rate $H^2$ is not always proportional to the dark energy of the
frozen modes due to the contributions in $H^2$ from other forms of energy
densitites.$H^2$ is dominated by frozen modes (and thus proportional
to the dark energy $\rho_{DE}$) only at some late times $t \ge t_E$ when
all other energy contributions $\rho_{other}$ have diluted enough below
$\rho_{DE}$ due to their redshift.

The quantitative effort we have made in this work suggests 
that an interpretation of the dark energy 
in terms of string theory is more convincing than either
a simple cosmological constant or the use of a slowly-
varying scalar field with fine tuned parameters.

\section*{Acknowledgements} 

PHF acknowledges the hospitality of Scuola Normale Superiore, Pisa,
where some of this 
work was done and the support of the Office of High Energy, US Department
of Energy under Grant No. DE-FG02-97ER41036 and thanks
R. Brandenberger and R. Rohm for comments on the 
manuscript  and several
participants of the M-theory Cosmology Conference at DAMTP, CMS,
Cambridge for useful discussions.
LM is grateful to L. Kofman, A. Peet, R. Bond, B. L. Hu, A. Tseytlin,
L. Parker and A. Kempf for beneficial discussions and would like to
thank CITA for its hospitality where part of this work was done.

\bigskip

\newpage

\bigskip
\bigskip
\bigskip

\noindent {\bf Notes Added.}

\bigskip
\bigskip

\noindent (i) We have not addressed the equation of state $w = p/\rho$
for dark energy in this work.  A recent paper by Lemoine {\it et al.}
({\tt hep-th/0109128}) attempts to calculate $w$ from a dispersion relation
similar to that discussed in the present paper,
arrives at a bizarre result $w = -186$ and concludes correctly that this
value appears in contradiction with data. However, this result for $w$ depends
sensitively on the assumed initial conditions ({\it e.g.} Eq. (34) of that
work). We agree with Lemoine {\it et al.} that an acceptable equation of
state is not automatic but depends
on initial conditions. Our conclusion would be that an acceptable equation of state
requires initial conditions dictated by string theory.  We hope 
to return to this question in a future
publication.

\bigskip
\bigskip

\noindent (ii). Starobinsky ({\tt astro-ph/0104043}) suggests that
trans-Planckian physics is irrelevant to inflationary cosmology;
however, his result is based on a 
discussion using the WKB approximation and as he himself emphasizes
does not apply to a dispersion relation which falls to
zero, $\omega(k) \rightarrow 0, k \rightarrow \infty$,
at large $k$ as we have assumed in the present article.

\newpage

\section*{Appendix}
\appendix

In this appendix we summarise the derivation of the coarse-grained
action $S_{eff}[\Psi_S]$, Eq. (\ref{action}). For details we refer the
reader to the original papers in \cite{blhu}. 

We start with the euclidean action for the quartic interaction scalar
model under consideration, Eq. (\ref{hamiltonian}), in a
spatially-flat FRW universe,
\begin{equation} 
S[\Psi] = \int a(t)dt \int d^3x \left[ |\nabla
\Psi|^2 + \tilde{m}^2_0 
|\Psi|^2 + \tilde{g}_1 
|\Psi|^4 \right]\,.  
\end{equation}
Time in this approach is considered as a parameter, and 
the scale factor $a(t)$ is regarded as a constant instead of a
dynamical function. In this way, different values of $a(t)$=constant
labels different spatial sections in 3-dimensional Minkowsky space,
related to each other by ``scaling'' transformations, with scaling
parameter 
\begin{equation}
b=a(t)/a(t_0) \,. 
\end{equation}
In this sense, $dynamics$ is replaced by $scaling$.    

In (3-dimensional) momentum space, the action is given by:
\begin{eqnarray} 
S[\Psi] &=& \int a(t) dt \left[ \int d^3p \Psi(p) \left(p^2 + \tilde{m}_0^2 \right)
\Psi(-p) \right. \nonumber \\
& & + \tilde{g}_1 \left. \int \prod_i^4 d^3 p_i
\Psi[p_1] \Psi[p_2] 
\Psi[p_3] \Psi[p_4] \delta^3(p_1+p_2+p_3+p_4) \right] \,.
\end{eqnarray} 
We now separate the field modes into system and environment, $\Psi=
\Psi_S + \Psi_E$, with the choice:
\begin{equation}
\Psi_S : \;p^S \leq \Lambda/b \,,\;\;\;\;\;
\Psi_E : \; \Lambda/b \leq p^E \leq \Lambda \,. 
\end{equation} 
We can then also write the action as system+environment, plus the
interaction between system and environment: 
\begin{equation}
S[\Psi]=
S[\Psi_S]+S_0[\Psi_E]+S_I[\Psi_E,\Psi_S]\,, 
\end{equation} 
where:
\begin{eqnarray} 
S_S[\Psi_S] &=& \int a(t)dt \left[ \int d^3 p^S \Psi_S(p^S) G_S^{-1}
\Psi_S(-p^S)  + \tilde{g}_1
\int \prod_{i=1}^{4} d^3 p^S_i \Psi_S^4(p^S_i) \delta^3(\sum p^S_i)
\right]\,, \\  
S_0[\Psi_E] &=& \int a(t)dt \int d^3 p^E \Psi_E(p^E) G_E^{-1}
\Psi_E(-p^E) \,, \\ 
S_I[\Psi_E,\Psi_S] &=& \tilde{g}_1 \int a(t) dt \left[ \int
\prod_{i=1}^4 d^3 p_i^E  
\Psi_E(p_1^E)\Psi_E(p_2^E)\Psi_E(p_3^E)\Psi_E(p_4^E) \delta^3(\sum
p_i^E) \right. \nonumber \\ 
&+& 4 ~\int \prod_{i=1}^3d^3p_i^E~ d^3p^S
\Psi_E(p_1^E)\Psi_E(p_2^E)\Psi_E(p_3^E)\Psi_S(p^S)
\delta^3(p^E_1+p^E_2+p^E_3+p^S) \nonumber \\
&+& 6 ~\int \prod_{i=1}^2d^3p_i^E~ \prod_{j=1}^2d^3p_j^S
\Psi_E(p_1^E)\Psi_E(p_2^E)\Psi_S(p_1^S)\Psi_S(p_2^S)
\delta^3(p^E_1+p^E_2+p^S_1+p_2^S) \nonumber \\
&+& 4~\left. \int d^3p^E~ \prod_{j=1}^3d^3p_j^S
\Psi_E(p^E)\Psi_S(p_1^S)\Psi_S(p_2^S)\Psi_S(p_3^S)
\delta^3(p^E+p^S_1+p^S_2+p_3^S) \right]\,, \\
\end{eqnarray}
where $G_{S,E}$ are Green's functions given by:
\begin{eqnarray}
G_S^{<\Lambda/b}[p^S< \Lambda/b]&=& [ (p^S)^2 + \tilde{m}_0^2]^{-1} \,, \\
G_E^{\Lambda/b}[p^E > \Lambda/b]&=& [ (p^E)^2 + \tilde{m}_0^2]^{-1} \,,
\end{eqnarray}
and the Green's function for the whole system is defined as
\begin{equation}
G[p]= G_S^{<\Lambda/b}[p^S] + G_E^{\Lambda/b}[p^E]\,.
\end{equation}
Now, the ``environment'' fields can be integrated out from the
partition function, such that:
\begin{eqnarray} 
Z[\Psi] &=& N \int D \Psi e^{-S[\Psi]} = N \int D \Psi_S \int D \Psi_E
e^{ -(S[\Psi_S] + S_0[\Psi_E] +  S_I[\Psi_E, \Psi_S])} \\
&=& N^\prime \int D \Psi_S e^{ -S[\Psi_S]} \langle e^{-S_I[\Psi_E,
\Psi_S]} \rangle_{\Psi_E} \,.
\end{eqnarray} 
The average $\langle \cdots \rangle_{\Psi_E}$ is defined with respect
to the free action for the environment fields $S_0[\Psi_E]$, 
\begin{eqnarray} 
\langle e^{-S_I[\Psi_E,
\Psi_S]} \rangle_{\Psi_E} &=& \int D\Psi_E
e^{-(S_0[\Psi_E]+S_I[\Psi_E,\Psi_S])}/ \int D\Psi_E  e^{-S_0[\Psi_E]}
\\
&=& e^{-\Delta S[\Psi_S]} \,.
\end{eqnarray} 
The coarse-grained effective action is then
\begin{equation}
S_{eff}[\Psi_S] = S[\Psi_S] + \Delta S[\Psi_S] \,,
\end{equation}
with 
\begin{equation}
\Delta S[\Psi_S] = - \ln \langle exp~ S_I[\Psi_E,\Psi_S] \rangle \,. 
\label{deltas}
\end{equation} 
When the interaction between system and environment modes is small
($\tilde{g}_1 \ll 1$), the contribution from Eq. (\ref{deltas}) can be
expanded in a Dyson-Feynman series, which contains only even powers of
the system fields (odd powers of the environment fields average to
zero). The first terms in the series give rise then to corrections to
the mass ($\delta \tilde{m}_0^2(b)$) and coupling constant ($\delta
\tilde{g}_1(b)$) parameters, which 
can be absorbed by redefining the original mass and coupling parameters in the
action $S_S[\Psi_S]$. Therefore, the effective action will be given
by:
\begin{eqnarray} 
S_{eff}[\Psi] &=& \int a(t) dt \left[ \int d^3p^S \Psi_S(p^S)
\left((p^S)^2 + \tilde{m}^2(b) \right)
\Psi_S(-p^S) \right. \nonumber \\
& & + \left. \int \prod_i^4 d^3 p^S_i
\tilde{g}(b) \Psi_S[p^S_1] \Psi_S[p^S_2] 
\Psi_S[p^S_3] \Psi_S[p^S_4] \delta^3(p^S_1+p^S_2+p^S_3+p^S_4) \right] \,,
\end{eqnarray} 
where
\begin{eqnarray}
\tilde{m}^2(b)&=& \tilde{m}_0^2 + \delta \tilde{m}_0^2(b) \,,\\
\tilde{g}(b)= &=& \tilde{g}_1 + \delta \tilde{g}_1(b) \,. 
\end{eqnarray}

\end{document}